\documentclass[aps,prd,superscriptaddress,twoside,twocolumn,nofootinbib,preprintnumbers]{revtex4}

\usepackage{amsmath}
\usepackage{graphicx}

\newcommand{\lesssim}{\mathrel{\mathpalette\vereq<}}

\begin{document}

\title{750 GeV Diphoton Signal from One-Family Walking Technipion}

\author{Shinya Matsuzaki}\thanks{{\tt synya@hken.phys.nagoya-u.ac.jp}}
      \affiliation{ Institute for Advanced Research, Nagoya University, Nagoya 464-8602, Japan.}
      \affiliation{ Department of Physics, Nagoya University, Nagoya 464-8602, Japan.}    
\author{Koichi Yamawaki} \thanks{{\tt yamawaki@kmi.nagoya-u.ac.jp}}
      \affiliation{ Kobayashi-Maskawa Institute for the Origin of Particles and the Universe (KMI) \\ 
 Nagoya University, Nagoya 464-8602, Japan.}

\date{\today}

\begin{abstract} 
The ATLAS and CMS groups have recently reported an excess 
at around 750 GeV with the local significance by about 3 sigma   
in the diphoton channel at the 13 TeV LHC. 
We give a possible explanation for the excess 
by a composite pseudo scalar $(P^0)$ predicted in 
the one-family model of walking technicolor.

\end{abstract}
\maketitle


Very recently, 
an excess about 3 sigma (at local significance) 
has  been seen at around 750 GeV in the diphoton mass distribution 
at the 13 TeV LHC experiments~\cite{ATLAS-750GeV,ATLAS-750GeV-2,CMS-750GeV,CMS:2015dxe}. 
It may indicate the existence of a new particle 
beyond the standard model.

In this paper, we present a  
possible explanation for 
the diphoton excess by a composite pseudoscalar boson, a pseudo Nambu-Goldstone boson of the chiral symmetry (technipion $P^0$) predicted in the TeV region 
in the one-family walking technicolor model~\cite{Jia:2012kd,Kurachi:2014xla}, 
a scale-invariant version~\cite{Yamawaki:1985zg,Bando:1987we} of the one-family technicolor model~\cite{Farhi:1980xs}, 
which successfully accounts for the LHC Higgs with the 125 GeV mass~\cite{Matsuzaki:2012xx,Matsuzaki:2015sya,Yamawaki:2015tmu} and the 
couplings~\cite{Matsuzaki:2012gd,Matsuzaki:2012vc,Matsuzaki:2012mk,Matsuzaki:2013fqa,Matsuzaki:2015pda}.

The model includes  
eight flavors $N_F=8$ of technifermions (techniquarks $Q_c$ and technileptons $L$) are introduced: 
$Q_c\equiv (U_c, D_c)^T$ 
(with $c=r,g,b$ being the QCD color charge) 
and $L \equiv (N, E)^T$, all having the technicolor $N_C$ of    
$SU(N_C)$~\cite{Farhi:1980xs}. 
The chiral symmetry is enhanced from that of 
the standard model $(SU(2)_L \times SU(2)_R)$ to 
$SU(8)_L \times SU(8)_R$, 
which is broken by the technifermion condensation 
$\langle \bar{F}F \rangle \neq 0$ ($F=Q, L$) down to $SU(8)_V$. 
One thus finds 63 composite pseudoscalar Nambu-Goldstone bosons ($\sim \bar{F} i \gamma_5 T^A F$, 
with $T^A$ ($A=1,\cdots, 63$) being the $SU(8)$ generators).   
Among 63, three are eaten by $W$ and $Z$ bosons, 
while other 60 become composite pseudo Nambu-Goldstone bosons  
(technipions) acquiring mass by  
the interactions outside of the technicolor sector, such as 
the extended technicolor and the standard-model gauge interactions,  
which break the chiral $SU(8)_L \times SU(8)_R$ symmetry 
in a way to keep only three exact 
Nambu-Goldstone bosons massless to be absorbed into $W$ and $Z$ 
while all others are massive.  
The masses are actually lifted up to be on the order of ${\cal O}(1)$TeV~\cite{Jia:2012kd,Kurachi:2014xla}, 
due to the large anomalous dimension $\gamma_m \simeq 1$ of the  
walking dynamics, 
a salient feature of the walking technicolor~\cite{Yamawaki:1985zg,Bando:1987we}.

In addition to the technipions, the walking technicolor possesses 
a light flavor-singlet scalar $(\sim \bar{F}F)$, technidilaton,
arising as a composite pseudo 
Nambu-Goldstone boson for the spontaneous 
breaking of the (approximate) scale invariance~\cite{Yamawaki:1985zg,Bando:1987we,Bando:1986bg}. 
It was shown to have mass  
as small as 125 GeV~\cite{Matsuzaki:2012xx,Matsuzaki:2015sya,Yamawaki:2015tmu} 
due to the walking nature characterized by the conformal phase 
transition~\cite{Miransky:1996pd}, 
particularly near the anti-Veneziano limit 
$N_C \rightarrow \infty$ with $N_C \alpha, N_F/N_C=$ fixed ($N_F/N_C\gg 1$)~\cite{Matsuzaki:2015sya,Yamawaki:2015tmu}, 
in contrast to the original technicolor of 
naive QCD scale up,   
identified as the LHC 125 GeV Higgs. 
Such a light flavor-singlet scalar was also observed on the lattice for the large $N_F$ QCD with 
$N_F=8$~\cite{Aoki:2014oha,Aoki:2013qxa,LSD} 
as a concrete model of the one-family walking technicolor 
as well as $N_F=12$~\cite{Aoki:2013zsa,Fodor:2014pqa,Brower:2014ita}. 
It has been shown that 
the technidilaton has the coupling property 
consistent with the current LHC Higgs data~\cite{Matsuzaki:2012gd,Matsuzaki:2012vc,Matsuzaki:2012mk,Matsuzaki:2012xx,Matsuzaki:2015sya}.

The one-family walking technicolor predicts further rich composite spectra:
besides the pseudo Nambu-Goldstone  
bosons (technipions and technidilaton), 
the model predicts vector mesons ($\sim \bar{F} \gamma_\mu T^A F$, 
technirhos), having the mass around a few TeV. 
Recently, it has been shown~\cite{Fukano:2015hga,Fukano:2015uga,Fukano:2015zua} 
that the one-family walking technirhos can account for 
the diboson excess at around 2 TeV reported by   
the ATLAS collaboration at the 8 TeV LHC~\cite{Aad:2015owa}, 
consistently with the electroweak precision tests as well as 
the direct search limits from the LHC experiments.

Thus, the one-family walking technicolor has been becoming a
viable candidate 
not only on the field theoretical ground, 
but also from the phenomenological aspect tested at the LHC. 
In this paper, we shall give yet another 
evidence 
of the one-family walking technicolor: 
that is the 750 GeV, iso- and color-singlet technipion $(P^0)$ 
signature in the diphoton channel. 
It will be shown that the $P^0$-diphoton signal can explain 
the excess about 3 sigma recently reported from 
the 13 TeV LHC experiments.

The iso- and color-singlet technipion $P^0$, 
is constructed from one-family technifermions as 
$\sim 1/(4 \sqrt{3}) \left( 
\bar{Q} i\gamma_5 Q - 3 \bar{L} i \gamma_5 L \right)$~\cite{Farhi:1980xs}. 
As noted in Ref.~\cite{Jia:2012kd},
the $P^0$ couplings to the standard model particles arise from the non-Abelian 
anomaly of the chiral $SU(8)_L \times SU(8)_R$ gauged by the standard model charges. 
The coupling form can unambiguously be fixed by 
the Wess-Zumino-Witten construction~\cite{Wess:1971yu,Witten:1983tw} in terms of the chiral Lagrangian 
just like the case of QCD pions~\cite{Jia:2012kd}: 
\begin{eqnarray} 
S_{P^0 g g} 
&=&  - \frac{N_{C}}{16 \sqrt{3} \pi^2} \frac{g_s^2}{F_\pi} \int_{M^4}  \sum_{a=1}^8 
P^0 d G^a d G^a 
\,, \nonumber \\ 
S_{P^0 \gamma \gamma} 
&=&  \frac{N_{C}}{12 \sqrt{3} \pi^2} \frac{e^2}{F_\pi} \int_{M^4}   P^0 d A d A 
\,, \nonumber \\ 
S_{P^0 Z \gamma} 
&=&  \frac{N_{C}}{6 \sqrt{3} \pi^2} \frac{e^2 s_W}{c_W F_\pi} \int_{M^4}   P^0 d Z d A 
\,, \nonumber \\ 
S_{P^0 Z Z} 
&=&  \frac{N_{C}}{12 \sqrt{3} \pi^2} \frac{e^2 s_W^2}{c_W^2 F_\pi} \int_{M^4}   P^0 d Z d Z 
\,, \nonumber \\ 
S_{P^0 W W} 
&=& 
0 
\,, \label{P0-gauge-int} 
\end{eqnarray} 
where things have been written in terms of differential forms 
on the Minkowski-space manifold $M^4$, 
and $e, s_W, g_s$ and $F_\pi$ respectively 
denote the electromagnetic coupling, weak mixing angle $(s_W \equiv \sin\theta_W, 
c_W^2\equiv 1- s_W^2)$, the QCD gauge coupling,  
and the technipion decay constant fixed by the electroweak scale $v_{\rm EW}=246$ GeV 
as $F_\pi = v_{\rm EW}/\sqrt{N_D}$, with $N_D=N_{F}/2=4$ for the one-family model. 
Note that the $P^0$-$W$-$W$ coupling vanishes because of the cancellation 
 between techni-quark and -lepton contributions~\cite{Jia:2012kd}. 
The partial decay widths are computed 
to be 
\begin{eqnarray} 
\Gamma (P^0 \to gg) 
&=&
\frac{N_C^2 \alpha_s^2 G_F m^3_{P^0}}{12 \sqrt{2} \pi^3} 
\,, \nonumber \\ 
\Gamma (P^0 \to \gamma\gamma) 
&=&
\frac{N_C^2 \alpha_{\rm em}^2 G_F m^3_{P^0}}{54 \sqrt{2} \pi^3} 
\,, \nonumber \\ 
\Gamma (P^0 \to Z \gamma) 
& = &
\frac{N_C^2 \alpha_{\rm em}^2 G_F m^3_{P^0}s_W^2}{27 \sqrt{2} \pi^3 c_W^2} 
\left( 1- \frac{m_Z^2}{m_{P_0}^2} \right)^3 
\,, \nonumber \\ 
\Gamma (P^0 \to Z Z) 
& = &
\frac{N_C^2 \alpha_{\rm em}^2 G_F m^3_{P^0}s_W^4}{54 \sqrt{2} \pi^3 c_W^4} 
\left( 1- \frac{4 m_Z^2}{m_{P_0}^2} \right)^{3/2} 
\,, \nonumber \\ 
\Gamma (P^0 \to WW) 
&=& 
0 \,, 
\end{eqnarray} 
where $\alpha_{\rm em}\equiv e^2/(4\pi)$, $\alpha_s\equiv g_s^2/(4\pi)$ and 
use has been made of $1/v_{\rm EW}^2=\sqrt{2} G_F$ with $G_F$ being the Fermi constant. 
  Note that all the partial decay widths are proportional to $N_C^2$, 
so the branching ratios are independent of the number of technicolor $N_C$.

\begin{table}
\begin{tabular}{|c|c|c|}
\hline 
\hspace{20pt}  
$N_C$ 
\hspace{20pt}  
&
\hspace{5pt}   
3  
\hspace{5pt}  
&
\hspace{5pt}   
4 
\hspace{5pt} 
\\  
\hline \hline 
$\Gamma_{\rm tot}[{\rm GeV}]$ & 1.2 & 2.1 \\ 
\hline 
Br($P^0 \to gg$)[\%]  & 99.8  & 99.8 \\ 
Br($P^0 \to \gamma\gamma$)[\%] & $9.7 \times 10^{-2}$ & $9.7 \times 10^{-2}$ \\ 
Br($P^0 \to Z \gamma$)[\%]& $5.3 \times 10^{-2}$& $5.3 \times 10^{-2}$ \\ 
Br($P^0 \to ZZ$)[\%]& $7.3 \times 10^{-3}$ & $7.3 \times 10^{-3}$ \\  
\hline 
\end{tabular}
\caption{ 
The total width and branching fraction of the $P^0$ at 750 GeV 
in the one-family walking technicolor with $N_C=3$ and $4$. 
}
\label{tab:P0:decay}
\end{table}

Using the experimental values~\cite{Agashe:2014kda} 
$G_F \simeq 1.166 \times 10^{-5} \,{\rm GeV}^{-2}$, 
$\alpha_s \simeq 0.118$ (at the $Z$ mass scale), 
$s_W^2 \simeq 0.22$, 
$m_Z \simeq 91.2$ GeV, 
$\alpha_{\rm em} \simeq (128)^{-1}$ (at the $Z$ mass scale), 
one calculates the total width $(\Gamma_{\rm tot})$ and branching ratios (Br)  
by setting the $P^0$ mass to 750 GeV and choosing $N_C$  
to be a certain number, 
listed as in Table~\ref{tab:P0:decay}.  
The table shows that the $P^0$ is a very  narrow resonance with the width 
of ${\cal O}(1{\rm GeV})$, in accordance with the diphoton signal reported 
in Refs.~\cite{ATLAS-750GeV,ATLAS-750GeV-2,CMS-750GeV,CMS:2015dxe}, and almost perfectly couple to digluon, implying 
the large gluon-gluon fusion (ggF) cross section at the LHC.

Now we estimate the 750 GeV $P^0$ cross sections at the 13 TeV LHC 
produced through 
the ggF process to get 
\begin{eqnarray} 
\begin{array}{l|l|l} 
\hline 
\sigma_{\rm ggF}^{13\,{\rm TeV}}(P^0)[{\rm fb}]  & N_C= 3 & N_C=4 \\ 
\hline \hline 
gg &  
 7.7 \times 10^3&  1.4 \times 10^4 \\      
\gamma \gamma&  7.5&  13 \\ 
Z\gamma &  4.1& 7.3 \\      
ZZ &  5.6 \times 10^{-1}&  9.9 \times 10^{-1}  \\ 
\hline 
\end{array}
\,, \label{13TeV:cross}
\end{eqnarray}
\vspace{5pt}

\noindent 
where we have used the narrow width approximation with the parton distribution function
\texttt{CTEQ6L1}~\cite{Stump:2003yu}.  
The table in Eq.(\ref{13TeV:cross}) thus shows that 
the $P^0$ diphoton cross sections reach the amount enough to 
explain the 750 GeV diphoton excess, $\sigma \times {\rm Br} \sim 5 - 10$ fb 
read off from Refs.~\cite{ATLAS-750GeV,ATLAS-750GeV-2,CMS-750GeV,CMS:2015dxe}.

The 750 GeV $P^0$ signals should be consistent with the currently 
available LHC limits. 
 From Refs.~\cite{CMS:2015neg,Aad:2015mna,CMS:2015cwa,Aad:2014fha,Aad:2015kna} one can read off the 
95\% C.L. upper limits   
on scalar resonances with mass of 750 GeV at the 8 TeV LHC as 
\begin{eqnarray} 
 \sigma^{8\,{\rm TeV}} \times {\rm Br}(gg) [{\rm fb}] 
& \lesssim & 5\times 10^3  
\,, \nonumber \\ 
 \sigma^{8\,{\rm TeV}} \times {\rm Br}(\gamma \gamma) [{\rm fb}] 
& \lesssim & 2.0  
\,, \nonumber \\ 
 \sigma^{8\,{\rm TeV}} \times {\rm Br}(Z \gamma) [{\rm fb}] 
& \lesssim & 4.0  
\,, \nonumber \\  
 \sigma^{8\,{\rm TeV}} \times {\rm Br}(ZZ) [{\rm fb}] 
& \lesssim & 12  
\,.  
\end{eqnarray}   
To these bounds, 
the 8 TeV cross sections of the $P^0$ are:  
\begin{eqnarray} 
\begin{array}{l|l|l} 
\hline 
\sigma_{\rm ggF}^{8\,{\rm TeV}}(P^0)[{\rm fb}]  & N_C= 3 & N_C=4 \\ 
\hline \hline 
gg &  
 1.7 \times 10^3&  3.0 \times 10^3 \\      
\gamma \gamma&  1.6&  2.9 \\ 
Z\gamma &  8.9 \times 10^{-1} & 1.6 \\      
ZZ &  1.2 \times 10^{-1}&  2.2 \times 10^{-1}  \\ 
\hline 
\end{array}
\,, \label{8TeV:cross}
\end{eqnarray}
\vspace{5pt}

\noindent 
which tells us that the $N_C=4$ case is in tension with  
the 8 TeV diphoton bound. 
Thus, one may conclude that 
the 750 GeV $P^0$ in the one-family walking technicolor with $N_C=3$ can 
account most favorably for the presently observed diphoton excess.

In conclusion, 
the iso- and color-singlet technipion $P^0$ in the one-family model of 
walking techniclor can have the diphoton signal consistent with 
the 13 TeV excess recently reported from the ATLAS and CMS groups. 
This would be a strong hint for the one-family walking technicolor 
to be evident in the real world, in addition to the 2 TeV walking technirho 
in accord with the diboson excesses. 
If the $P^0$ is present in the diphoton channel, 
then some excesses in other channels expected from the numbers listed in 
Eq.(\ref{13TeV:cross}), such as 
in dijets, $Z\gamma$ and $ZZ$ channels, would promisingly 
be seen in the near future LHC Run-II data. 
More detailed study on the $P^0$ signatures in other channels 
and distinct signals from other walking technipions, 
such as QCD-colored ones,  
will be pursued elsewhere.

In closing, 
the $P^0$ could couple to the standard model fermions 
through extended technicolor interactions,   
as discussed in Ref.~\cite{Jia:2012kd}, although they are formally higher loops. 
Among the standard model fermions, 
the Yukawa coupling to top quark pair would be most influential with either constructive or attractive interference with the Wess-Zumino-Witten 
term to give significant corrections to 
the branching fraction of the $P^0$, including possible relaxing the 8 TeV LHC constraints. 
Also, Yukawa couplings might be  
constrained by 
the possible excessive flavor-changing neutral current processes.  
Since such Yukawa coupling forms are highly model-dependent on 
details of the extended technicolor model building,    
this issue deserves to another publication in the future.

\acknowledgments

We thank Masaharu Tanabashi for valuable discussions and encouragements. 
This work was supported in part by 
the JSPS Grant-in-Aid for Young Scientists (B) \#15K17645 (S.M.).  
\\

\noindent 
{\it Note added}\\ 
\noindent 
After having finished the paper, 
we noticed a paper, arXiv.1512.05334,
discussing the 750 GeV diboson excess
in the technicolor framework. In contrast to
their ``$\eta$''-like  pseudo-scalar for $N_F=2$, our technipion in the one-family walking technicolor has 
enough production cross section due to the colored technifermions (techniquarks), and has no $WW$ coupling.

\end{document}